\documentclass{sig-alternate}

\usepackage[utf8]{inputenc}
\usepackage{wrapfig}
\usepackage{cite}
\usepackage{graphicx}
\usepackage{caption}
\usepackage{subcaption}
\usepackage[numbers]{natbib}
\usepackage{fancyhdr}
\usepackage{float}
\usepackage{fixltx2e}

\begin{document}

\title{Indian Premier League (IPL), Cricket, Online Social Media}
\numberofauthors{3} 
%
\author{
%
\alignauthor
Megha Arora
\thanks{Megha and Raghav did this work as a part of their summer work at Precog (http://precog.iiitd.edu.in/).}\\
       \affaddr{Indraprastha Institute of Information Technology - Delhi}\\
       \email{megha12059@iiitd.ac.in}
\alignauthor
Raghav Gupta\textsuperscript{ *}\\
       \affaddr{Carnegie Mellon University}\\
       \email{raghavg@andrew.cmu.edu}
\alignauthor Ponnurangam Kumaraguru\\
       \affaddr{Indraprastha Institute of Information Technology - Delhi}\\
       \email{pk@iiitd.ac.in}
}
\maketitle

\begin{abstract}
In recent past online social media has played a pivotal role in sharing of information and opinions on real time events. Events in physical space are reflected in digital world through online social networks. Event based studies for such content have been widely done on Twitter in the computer science community. In this report, we performed a detailed analysis of a sports event called  the Indian Premier League (IPL'13) for both Facebook and Twitter. IPL is the most popular cricket league in India with players from across the world. We analysed more than 2.6 million tweets and 700 thousand Facebook posts for temporal activity, text quality, geography of users and the spot-fixing scandal which came up during the league. We were able to draw strong correlations between the brand value of teams and how much they were talked about on social media across Facebook and Twitter. Analysis of geo-tagged data showed major activity from metropolitan suburbs however activity was not restricted to the regions geographically associated with each team. We present a decay calculation methodology, using which we derive that activity died down on both Twitter and Facebook in a very similar manner. Such analysis can be used to model events and study their penetration in social networks. We analysed text for spot-fixing and found that user response to allegations about matches being fixed was cold. The complete analysis presented in this report, can be particularly useful for studying events involving crisis or events of political importance having similar structure. 
\end{abstract}

\section{Introduction}

Over the past few years, there has been an increase in the usage of Online Social Media (OSM) services as a medium for people to talk and discuss events. Genres of these events include politics, science, sports, entertainment, culture, philosophy and literature. Significant events like political elections, riots, protests and crisis situations like predicted natural disasters and terrorist attacks have been widely discussed on online social platforms. Our motivation has been to analyze an event having a defined schedule and pre-scheduled sub-events, which involves stakeholders and their supporters, so that our analysis can be beneficial for such studies. Our analysis gives useful insights into the decay of events in the digital world of social networks.
\vspace{3 mm}

Sports events trend on all online social networks.\footnote{https://2013.twitter.com/\#category-sports} One such prolonged and popular event was the Indian Premier League (IPL). It is a showcase for Twenty-20 (T20) cricket, which is the shortest form of play in the game. Top Indian and international players participate in IPL, making it the world's richest cricket tournament.\footnote{http://www.bbc.co.uk/news/world-asia-india-19921111} IPL has been the subject of several controversies involving allegations of cricket betting, money laundering and spot fixing. The year 2013 marked the sixth season of IPL. The league started on April 3rd, 2013 and continued till May 26th, 2013. The nine teams which took part in the sixth season were Bangalore Royal Challengers, Chennai Super Kings, Mumbai Indians, Delhi Daredevils, Kings XI Punjab, Kolkata Knight Riders, Pune Warriors India, Sunrisers Hyderabad and Rajasthan Royals.\footnote{http://www.iplt20.com/} Every team name is associated with a city or state in India; like Rajasthan Royals is associated with the state of Rajasthan in the western part of the country, while Delhi Daredevils is associated with Delhi which is the national capital. However, players for each team came from different parts of the country and cricket teams from all across the world. All the teams competed against each other in 72 matches which were followed by two qualifiers, one eliminator and the finals. During these 76 matches, there were days with no matches, a single match or two matches.

\vspace{3 mm}
IPL’13 particularly became a huge hit in Online Social Media \cite{Sharma}.
Teams were ranked according to their presence on popular online social networks. Chennai Super Kings led with an online social footprint of over 2.49 million followed by Kolkata Knight Riders with 2.14 million \cite{Mishra}. Particular players and their performances were also a major topic of discussion throughout the league on OSNs. In this report, we analyze in detail the activity on Facebook and Twitter during the league. We attempt to study user behavior, amount of content, quality of content, user geography, team support and spot-fixing of the league in detail. We correlate the activity on Facebook and Twitter with the performance of each of the teams during the season, its brand value and the number of supporters. Analysis of the spot-fixing controversy gives an interesting insight into how a sub-event within IPL aligns with the league as a whole.
\vspace{3 mm}

IPL shares structural similarity with a wide variety of events. Our decay analysis for IPL has been tested to work well for events of all genres as shown in Section 4. Geo-analysis of events, as we have done for IPL in this report  with upcoming location based social networks like Foursquare\footnote{https://foursquare.com/} and Facebook Places. Unlike many event based analysis that have been done in the computer science community, our study is not restricted to Twitter. We have analyzed than 2.5 million tweets and 700 thousand Facebook posts in this study.

\section{Related Work}

Popularity of content on the Web, like news articles \cite{Szabo}, blog posts \cite{Mei, Kumar} and posts in online discussion forums \cite{Aperjis}, vary on different temporal scales. For example, content on micro-blogging platforms, like Twitter is very volatile, and pieces of content become popular and fade away in a matter of hours \cite{Java}. Blogging and micro-blogging networks show temporal and topological patterns which largely exhibit power law behavior \cite{Kwak, Leskovec}. Sitaram et al. have studied the growth and decay of trending topics on Twitter \cite{Asur}. Temporal trends displayed by users have been analysed and the decay factor has been found to decrease in a power law fashion. Yang and Leskovec examined patterns of temporal behavior for hashtags in Twitter \cite{Yang}. They presented a stable time series clustering algorithm and demonstrated the common temporal patterns that tweets containing hashtags follow.

\vspace{3 mm}
Many event-based studies have been done in the recent past. Kairam et al. studied trending events during peak activity periods \cite{Kairam}. Identification, prediction, classification and diffusion patterns of real world events have been widely analyzed (Kim et al. 2012 \cite{Kim}, Becker et al. 2010 \cite{Becker}, Abbasi et al. 2011 \cite{Abbasi}). Events of different genres have been covered in these analysis. Zin et al. presented a knowledge based event analysis framework for automatically analyzing key events by using various social network sources in case of disasters \cite{Zin}. Sakaki et al. investigated real time interaction of events like earthquakes on Twitter \cite{Sakaki}. Gupta et al. studied crisis situations and their content credibility for events like the Mumbai blasts \cite{Gupta} and the Boston Blasts (2013).

\vspace{3 mm}

To the best of our knowledge, this is the first attempt to study in detail, an India-centric sports event on online social media. Paridhi et al. studied the cross pollination of information on FIFA World Cup 2010 from Flickr, YouTube and Foursquare to Twitter \cite{Paridhi}. Burnap et al. examined data related to the London Olympics 2012 to conceptualize the relationship between social actors, events and social media for understanding the dynamic reactions of populations \cite{Burnap}. Clavio et al. have tried to identify the demographic characteristics of a sample of college football fans \cite{Clavio}. Kewalramani et al.have identified clusters in tweets for IPL 2011 \cite{Kewalramani}. They labeled the clusters based on the teams that were a part of the league.

\section{Data Collection}
We used MultiOSN \cite{Dewan} for data collection. MultiOSN monitors real world events on multiple online social media. The query terms have been listed in Table 6 (Appendix). This data had to be filtered because terms like `MI' which were actually used to collect data corresponding to Mumbai Indians, a team in IPL, may be used in other contexts as well such as referring to Michigan state or `mi' in Spanish. We filtered the data using a more accurate and exhaustive list of hashtags related to IPL and keywords consisting of abbreviations, names of teams, and players.
We collected data of over 4 million tweets from Twitter and over 3 million public posts from Facebook. Nearly 1.37\% of the tweets and 0.75\% of the Facebook public posts were geo-tagged (Table 1). Official statistics for IPL'13 claim that 6.7 million tweets were recorded from one million users over a period of two months. Similar data for Facebook is not available. Our filtered dataset for Twitter comprises of nearly 2.6 million tweets (around 39\% of the claimed data) from about 485 thousand unique users. 
 
\begin{table}[ht]
\begin{tabular}{l l l} 
\\[0.1ex]\hline\hline
&Twitter & Facebook \\ [0.3ex] 
\hline \\ 
Total Tweets/Posts  & 4,895,784 & 3,753,499 \\ [0.4ex] 
IPL Tweets/Posts (filtered)  & 2,627,197  & 774,186 \\ [0.4ex] 
IPL Unique users  & 485,533 & 401,254\\ [0.4ex] 
Geo-tagged Tweets/Posts  & 36,126 &  5,827\\ [0.4ex] 
Start Date (mm/dd/yy)  & 04/01/13 &  04/01/13 \\ [0.4ex] 
End Date (mm/dd/yy)  & 05/31/13 & 05/31/13 \\ [0.4ex] 
\hline 
\end{tabular}
\label{table:nonlin} 
\caption{Descriptive statistics of the data collected.}
\end{table}

We also analysed the quality of data for Twitter. Figure 1 represents the number of tweets per user for all the users who tweeted about IPL. We can clearly observe a power law curve for the number of tweets per user, which is in agreement with previous research work on Twitter \cite{Gupta, Kwak}.

\begin{figure}[h]
    \includegraphics[width=0.47\textwidth]{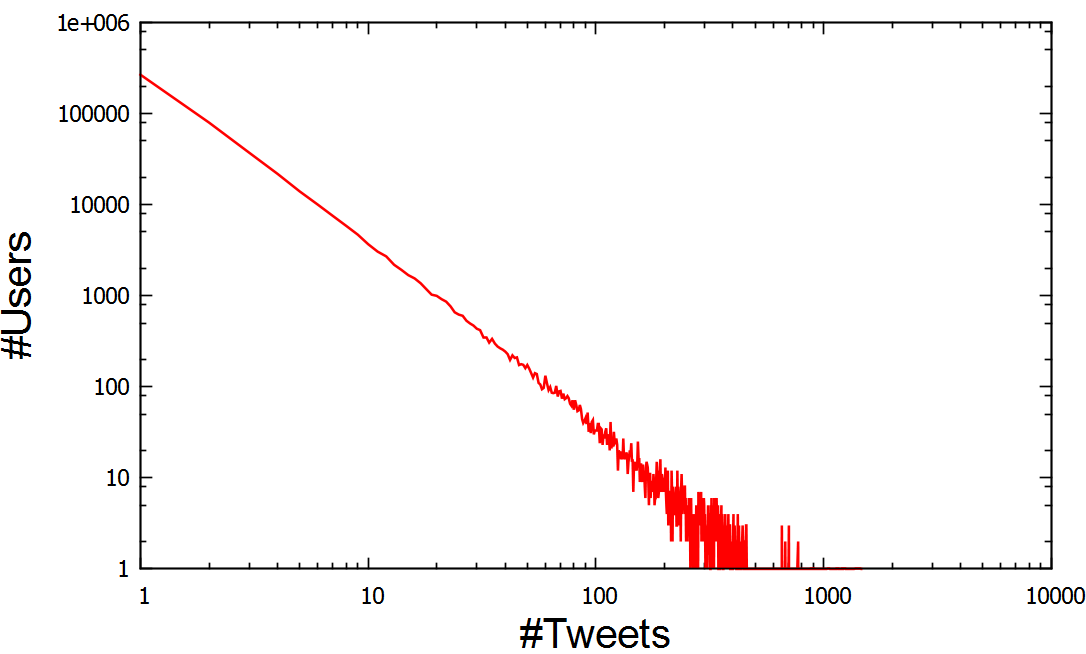}
  \caption{Shows the distribution of number of tweets per user on a logarithmic scale from the Twitter dataset.}
\end{figure}

\vspace{-1em}
\section{Decay Methodology}

The way in which the activity of an event dies down is indicative of the impact an event has. The decay can be steep for short-lived events while events which have an extended presence on OSNs depict a gentler fall. We analysed the decay in activity on Twitter for the Texas fertiliser plant explosion, Boston Marathon bombing, the Mumbai blasts and various other events. Figure 2 shows the fall in hourly tweet count and net decay factor for some of these events. We observed that the duration of decay may vary from a few hours to a few days. Another contributing factor can be an increase in the activity during the decay of an event. We can take into account all this information and calculate the net decay factor.

We obtained a graph for each event which depicted the activity per hour on the vertical axis and time on the horizontal axis. From the plot, we identified the decay region i.e. the region after the maximum activity. Starting from this peak value (marked with *  in each plot in Figure 2), we considered all the data points except when the activity reduced to less than 99\% of the peak activity. This was defined as the threshold for insignificant decay. Beyond this limit, the activity is not included in the decay factor calculation.
Within the decay, there can be some very steep peaks which we found could be linked to sub-events. In the event of a news update, there is a sudden surge in the activity but it dropped as quickly as it rose. These peaks do not correctly depict the actual decay of the event and need not be considered for the decay factor calculation. For our calculations we removed the peaks before recursively dividing the decay region into smaller regions. The divisions were made such that the R\textsuperscript{2} value for the fit of the line having an equation of the form of Equation (1) was more than 0.8.

\vspace{1 mm}

\begin{math}
\quad
\quad
\quad
\quad
\quad
\quad
\quad
\quad
\chi
\end{math}
=
\begin{math}
\alpha
\ln
(t)+
\beta
\quad
-
\quad
(1)
\end{math}

\vspace{1 mm}

where 
\begin{math}
\chi
\end{math}
represents activity, t represents time and
\begin{math}
\alpha
\end{math}
 and 
\begin{math}
\beta
\end{math}
are constants.

This particular equation has been chosen considering characteristics of plots for OSM activity and also by applying various possible exponential and logarithmic models. For each region the value
\begin{math}
\alpha/\beta
\end{math} is considered as its contribution to the net decay as it is directly proportional to the fall depicted in the region. Similarly for regions having a positive slope i.e. increase in activity, will have a negative 
\begin{math}
\alpha/\beta
\end{math}
 ratio. Calculations for the decay-growth pairs occurring simultaneously are then aggregated to represent the pair as one region instead of two. The values obtained from each region are then aggregated depending on the size of the region thus taking into consideration its contribution to (\begin{math}
\alpha/\beta
\end{math})$_{\text{net}}$.

\begin{figure}[h!]
        \begin{subfigure}[t]{0.5\textwidth}
                \centering
                \includegraphics[scale=0.283]{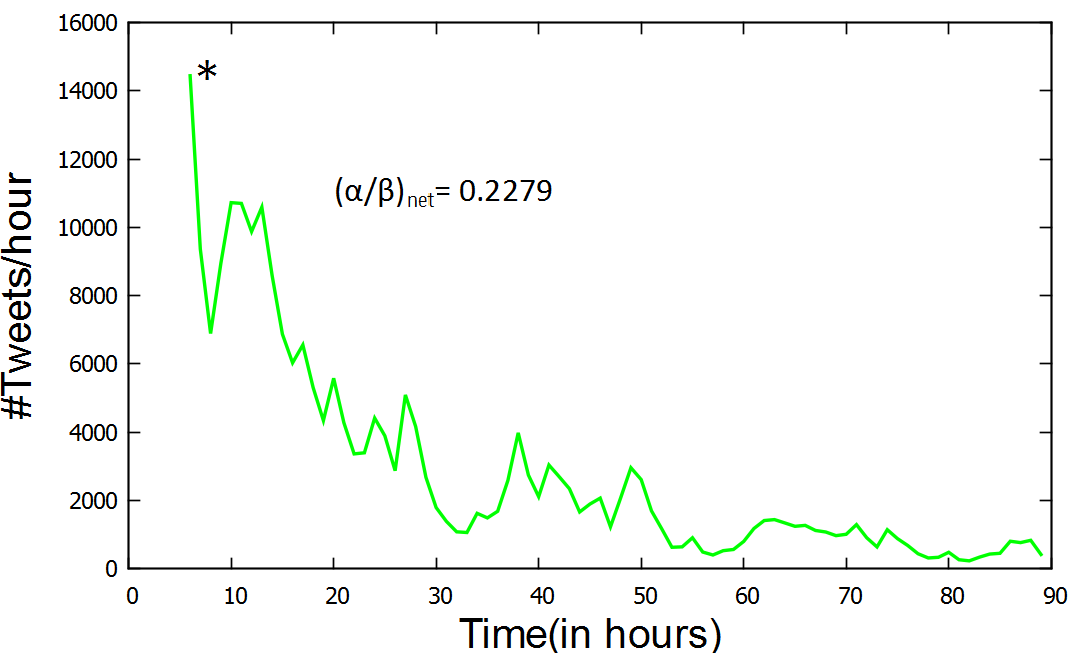}
                \caption{Fertiliser plant explosion, Texas}
                \label{fig:tiger}
        \end{subfigure}
        \begin{subfigure}[t]{0.5\textwidth}
                \centering
                \includegraphics[scale=0.283]{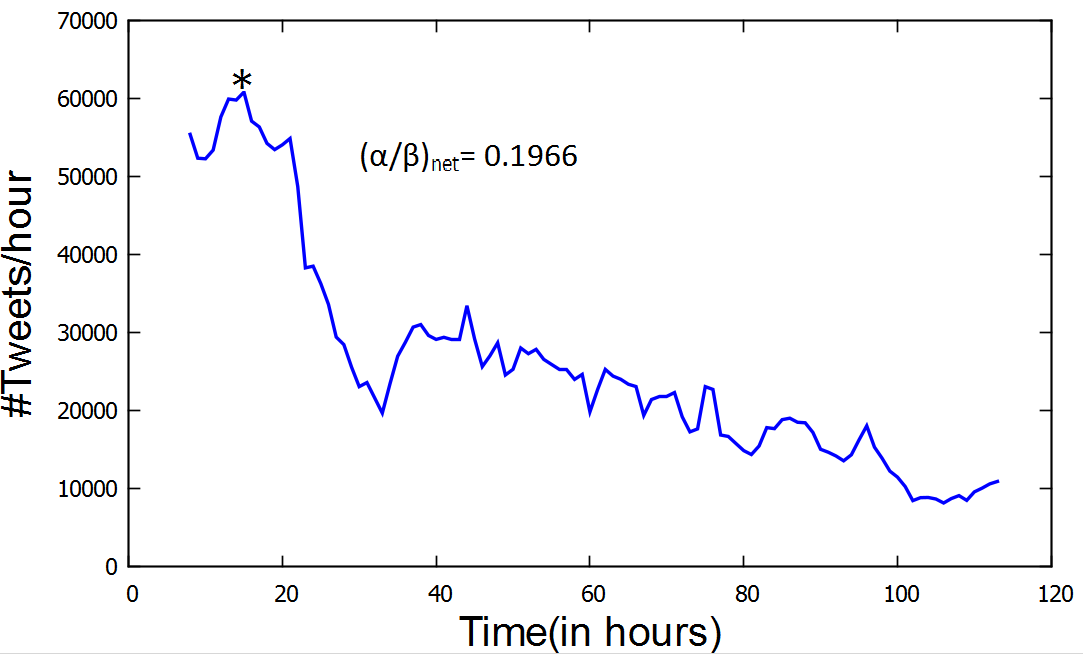}
                \caption{Boston Marathon Bombing}
                \label{fig:mouse}
        \end{subfigure}
        \begin{subfigure}[t]{0.5\textwidth}
                \centering
                \includegraphics[scale=0.283]{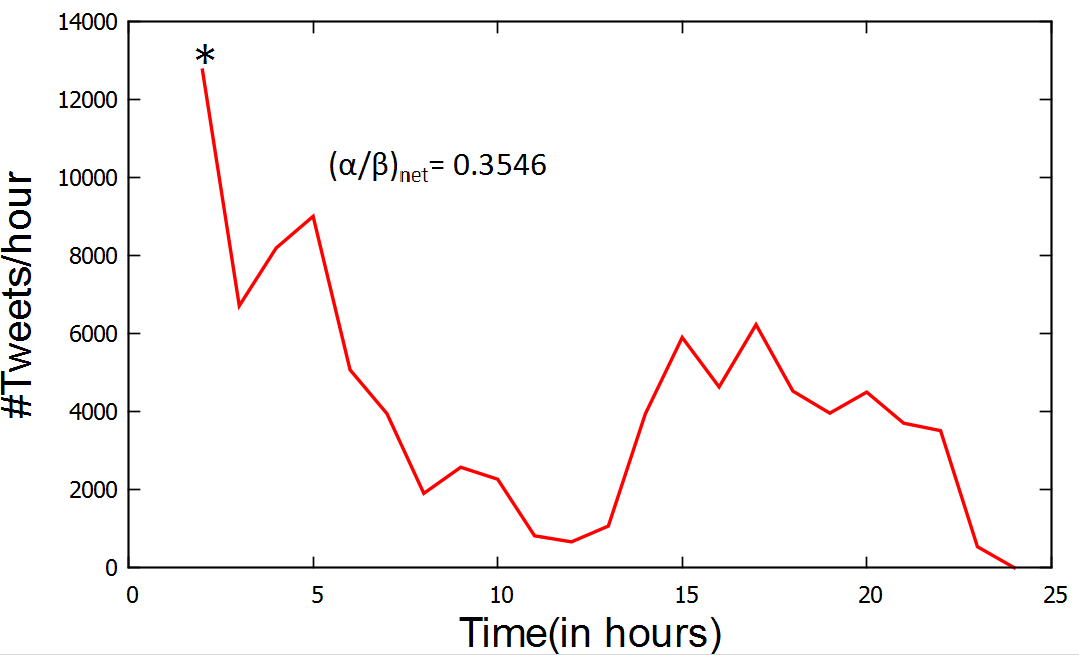}
                \caption{Mumbai Blasts}
                \label{fig:tiger}
        \end{subfigure}
	\caption{Decay in hourly tweet count along with net decay factor for Texas fertiliser plant explosion, Boston marathon bombing and Mumbai blasts.}\label{fig:animals}
\vspace{-5 mm}
\end{figure}
\vspace{-3 mm}

\section{Analysis}

The plots shown in Figures 3 represent data collected over two months during the IPL season. The peaks coincide with the different matches and the last peak coincides with the final match. These plots clearly indicate lesser data for Facebook as compared to Twitter. This may be attributed to the more public nature of Twitter as compared to Facebook and also to the absence of user data and non-public posts which cannot be obtained using the Facebook search API.\footnote{https://developers.facebook.com/docs/reference/api/} We also looked at day-wise plots and as shown in the plots from the Facebook dataset (Figure 4), these are clearly indicative of the number of matches that were played on a particular day. The plots obtained from the Twitter dataset share this property. We can infer from the plots that most activity has been recorded during the matches and not after or before.

\begin{figure*}[h!]
        \begin{subfigure}[t]{0.5\textwidth}
                \centering
                \includegraphics[scale=0.283]{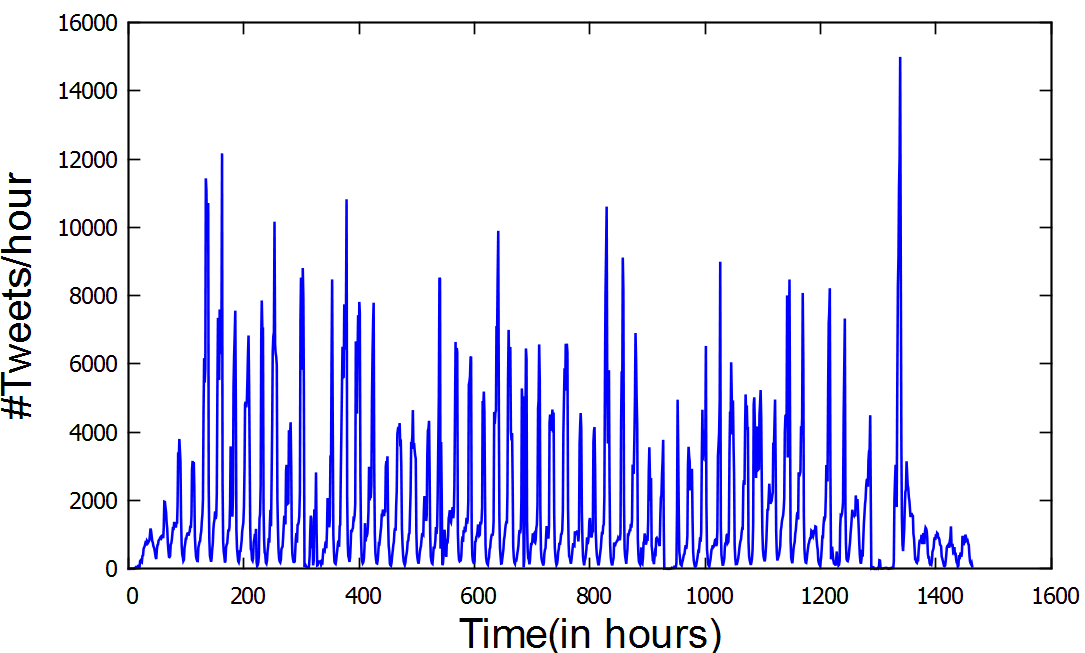}
                \caption{Tweets vs. Time}
                \label{fig:tiger}
        \end{subfigure}
        \begin{subfigure}[t]{0.5\textwidth}
                \centering
                \includegraphics[scale=0.283]{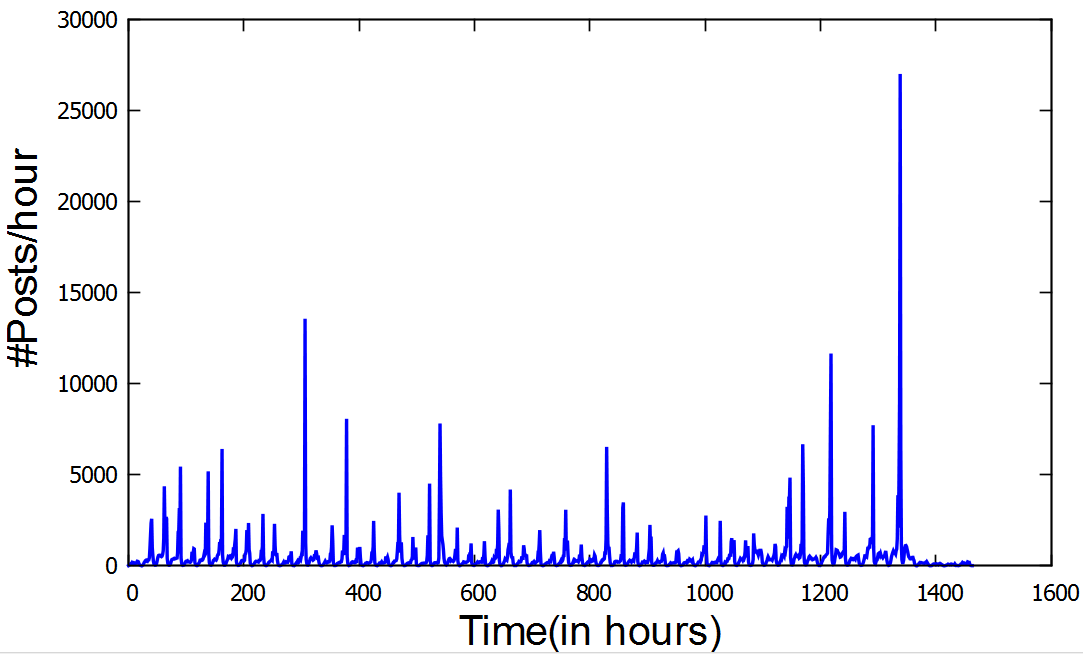}
                \caption{Facebook Posts vs. Time}
                \label{fig:mouse}
        \end{subfigure}
	\caption{Hourly data collected for Twitter and Facebook over entire IPL.}\label{fig:animals}
\end{figure*}

\begin{figure}[h!]
        \begin{subfigure}[t!]{0.5\textwidth}
                \centering
                \includegraphics[scale=0.283]{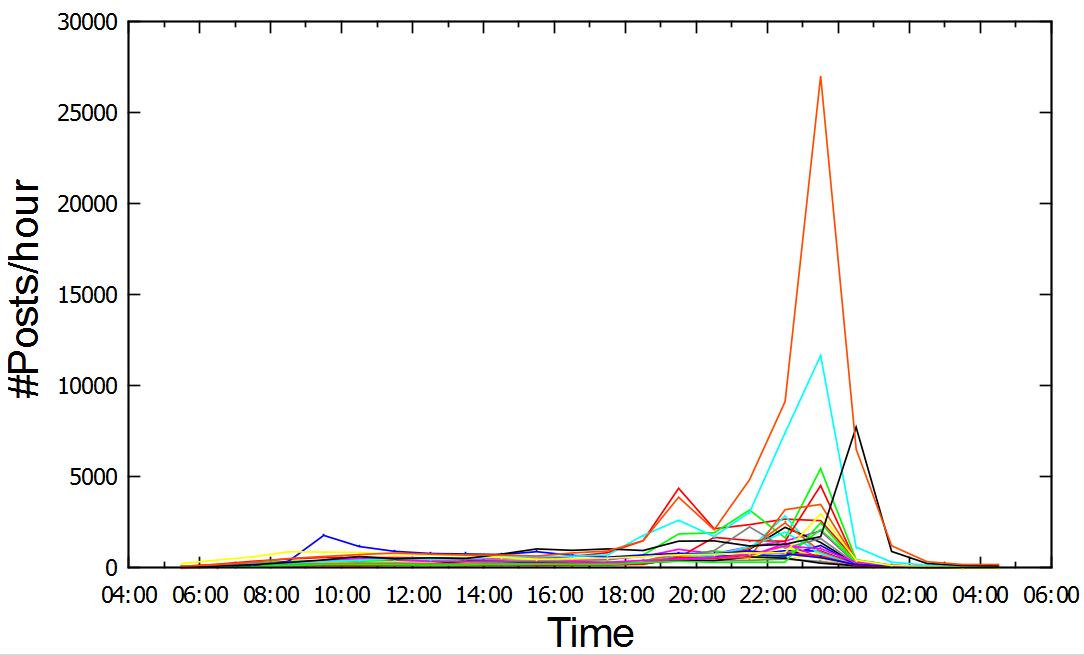}
                \caption{Activity on days having one match (total 26 days).}
                \label{fig:tiger}
        \end{subfigure}
        \begin{subfigure}[t!]{0.5\textwidth}
                \centering
                \includegraphics[scale=0.283]{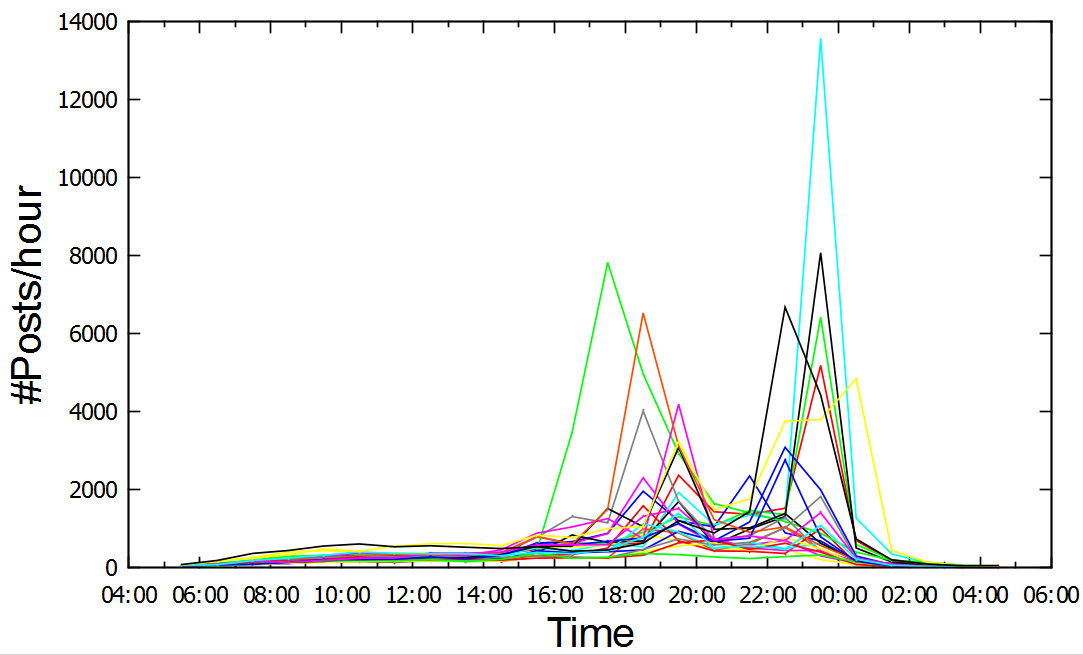}
                \caption{Activity on days having two matches (total 25 days).}
                \label{fig:mouse}
        \end{subfigure}
\vspace{1 mm}
        \caption{Daywise plots for Facebook data clearly depicting high frequency of posts during the match timings.}\label{fig:animals}
\end{figure}

\begin{figure}[h]
        \begin{subfigure}[h]{0.45\textwidth}
                \centering
                \includegraphics[scale=0.3]{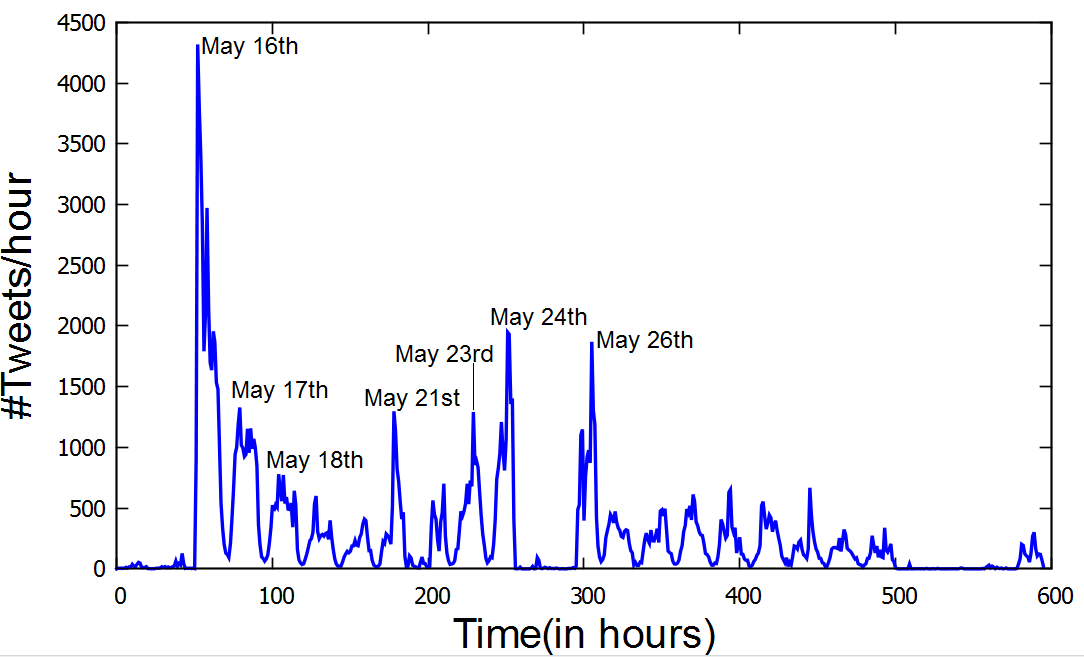}
                \caption{Twitter}
                \label{fig:tiger}
        \end{subfigure}
        \begin{subfigure}[h]{0.45\textwidth}
                \centering
                \includegraphics[scale=0.3]{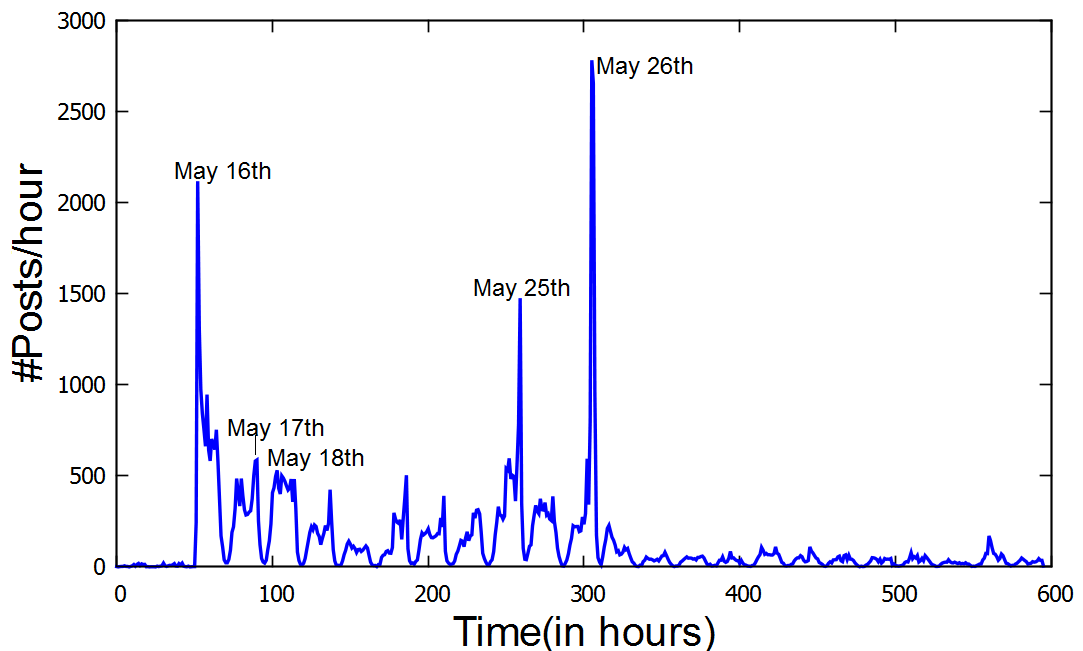}
                \caption{Facebook}
                \label{fig:mouse}
        \end{subfigure}
        \caption{ Plots from the spot-fixing dataset showing peaks in activity aligned with the spot-fixing timeline in Table 2.}\label{fig:animals}
\end{figure}

\vspace{2 mm}
A controversy arose during the season when three cricketers were arrested on charges of spot-fixing (dishonestly determining the outcome of a specific part of a game before it is played). Data for spot-fixing was filtered from the IPL Dateset starting 5:30 AM (IST) on the 14th of May till 12:30 AM (IST) on the 8th of June and activity was measured around specific keywords which  have been listed in Table 7 (Appendix). 167,503 tweets (10.3\% of the Twitter IPL data) and 80,312 Facebook posts (6.3\% of the Facebook IPL data) were obtained after filtering.  These statistics indicate that spot-fixing was one of the major sub-events under IPL. Plots for activity corresponding to spot-fixing yielded interesting results.

As can be seen in Figure 5 the frequency of peaks for the spot-fixing data is the same as the frequency of matches. We also found that in case of Twitter most peaks occur during match timings. The significance of the plots can be better understood using the spot fixing timeline given in Table 2.\footnote{http://www.dnaindia.com/sport/1844682/report-ipl-spot-fixing-controversy-a-timeline} The activity became significant soon after the three cricketers were arrested.We see that the peaks in the plots coincide with important advances and updates in the spot 
fixing controversy. We also observed that the activity on both platforms decays  significantly from 26th of May, the day on which the final match was played, officially marking an end to this season.
The spot-fixing controversy began only two weeks from the end of the IPL season, though its percentage in the total data indicates that spot-fixing contributed to a significant fraction of the activity, 14th of May onwards. On the day of the final match (26th of May), spot-fixing accounted for 14.1\% of the IPL activity on Twitter and 17.3\% of the same on Facebook. This gives us good insight into how sub-events work within events and how they can be analysed together. 

\vspace{-1 mm}

\begin{table}[ht]
\begin{center}
\begin{tabular}{l l l} 
\\[0.1ex]\hline
\hline \\
 May 16 & Rajasthan Royals players Sreesanth, \\ [0.1ex] 
& Ajit Chandila and Ankeet Chavan \\ [0.1ex] 
\vspace{1 mm}
 & arrested for spot-fixing by Delhi Police \\ [0.1ex] 

 May 17 &  BCCI suspends former Royals player,  \\ [0.1ex] 
\vspace{1 mm}
& Amit Singh, who was arrested as a bookie \\ [0.1ex] 

 May 18 &  Mumbai police link Ramesh, Ashok and \\ [0.1ex] 
 & Kadam to the case; Seize Sreesanth and \\ [0.1ex] 
 & Jiju Janardhan's belongings from  hotel \\ [0.1ex] 
&  rooms booked in their names at a \\ [0.1ex] 
\vspace{1 mm}
&  five-star hotel in Mumbai \\ [0.1ex] 

 May 21 & Vindoo Dara Singh arrested for alleged\\ [0.1ex]
\vspace{1 mm}
 & links with bookies \\ [0.1ex]

 May 23  & Mumbai police team raids Meiyappan's   \\ [0.1ex] 
& Chennai residence, summons Meiyappan for \\ [0.1ex] 
& questioning, rejects his request \\ [0.1ex] 
\vspace{1 mm}
& for an extension till after IPL-VI finals\\ [0.1ex] 

 May 24 &  Meiyappan arrives in Mumbai; He is \\ [0.1ex] 
\vspace{1 mm}
& arrested on charges of betting and cheating \\ [0.1ex] 

 May 25 & Mumbai police say Meiyappan passed on \\ [0.1ex] 
 &  information about Chennai Super Kings \\ [0.1ex] 
\vspace{1 mm}
& to bookies \\ [0.1ex] 

 May 26 & BCCI announces three-member commission  \\ [0.1ex] 
 & to investigate Meiyappan; ICC removes \\ [0.1ex] 
 & umpire Asad Rauf from the  Champions \\ [0.1ex] 
\vspace{1 mm}
 & Trophy squad\\ [0.1ex] 

\hline \hline 
\end{tabular} 
\caption{IPL spot-fixing timeline}
\end{center}
\end{table}
\vspace{-2 mm}

\subsection{Geo-Analysis}

We analysed the geo-tagged tweets and Facebook posts to better understand the geographic distribution of IPL activity on social media. Geo-tagged tweets only comprise a small percentage of the dataset as seen in Table 3. Even though the results are in coherence with what might be expected, the possibility of the geo-tagged dataset being biased cannot be ignored. We plotted heatmaps to understand the global distribution as well as the distribution within India. Apart from India which accounted for 75.5\% of the geo-tagged tweets (27,271) and 89.8\% of the Facebook posts (5,232) the list of other countries from where significant activity was seen included UK, UAE, USA, South Africa, Singapore, Indonesia, Sri Lanka and Pakistan. 

\vspace{3 mm}

Within India, the density is highest in the metropolitain areas. Delhi, Mumbai, Bangalore and Chennai are the cities with maximum activity as can be seen in Figure 6. Analysing addresses helped us further understand the distribution between cities and upcountry. In case of twitter we found that the top 50 most populated cities according to the census of India \footnote{http://censusindia.gov.in} accounted for 69.95\% of the tweets (19,075) and top 5 accounted for 46.98\% (12,812) of the tweets. For Facebook the figures were 81.82\% (4,281) and 55.22\% (2,889) respectively. 

\vspace{3 mm}

\begin{table}
\begin{center}
\begin{tabular}{l l l} 
\\[0.1ex]\hline\hline
&Twitter & Facebook \\ [0.1ex] 
\hline \\ 
Total Geo-Tagged Data  & 36,107 & 5,823 \\ [0.1ex] 
Percentage in total IPL Data(\%)  & 1.3 & 0.75\\ [0.1ex] 
\hline 
\end{tabular}
\caption{Geo-Tagged Dataset}
\end{center}
\end{table}

\begin{table}
\begin{tabular}{l l l l}
\hline\hline
Team & Geo Tweets  & Local activity \\ 
\hline \\ 
Chennai Super Kings  & 7025 & 18.90\% \\ [0.4ex] 
Mumbai Indians  & 5741 & 36.71\%  \\ [0.4ex] 
Royal Challengers  & 4730 & 16.74\% \\ [0.4ex] 
Pune Warriors & 1684 & 34.73\% \\ [0.4ex] 
Sunrisers Hyderabad & 2050  & 14.92\% \\ [0.4ex] 
Delhi Daredevils & 1869  & 18.56\% \\ [0.4ex] 
Kolkata Knight Riders & 2371 & 9.78\% \\ [0.4ex] 
Kings XI Punjab & 1449 & 4.07\% \\ [0.4ex] 
Rajasthan Royals & 2693 & 4.64\% \\ [0.4ex] 
\hline 
\end{tabular}
\label{table:nonlin} 
\caption{Team-wise description of geo-tagged dataset}
\end{table}

Upcountry refers to the remaining 30.05\% of the data i.e. tweets other than those from the 50 most populated cities in India.
We went on to study the distribution for the upcountry data for Twitter and found that most states and regions with high activity were among the more developed regions of the country and had high levels of literacy. The list of states having contributed most significantly included Maharastra, Karnataka, Tamil Nadu, Delhi, Kerela and Gujarat. There were also states with no geo-tagged contributions  on either of the social networks such as Jammu and Kashmir and Manipur.  Even in case of the upcountry data, a big chunk can be attributed to suburbs of big cities, locations along important highways and to industrial belts. Along with internet penetration, another contibuting factor to this skew is high prevalence of content in English, which is a hurdle for much of rural India.

\vspace{3 mm}

We also analysed heatmaps of the geo-tagged data for each team separately and though most of the activity came from the home state and cities of each team, there was a spread across the metropolitans for all the teams. So people have been talking not only of their home teams but also of teams from other parts of the country showing no regional or geographical bias. We see in Table 4 that while in the case of Mumbai Indians 36.71\% of the geo-tagged tweets were from Maharastra, for Kings XI Punjab only 4.07\% of the geo-tagged tweets were from Punjab. It is worth noting that both the teams from Maharastra, namely the Mumbai Indians and the Pune Warriors India have the highest contributions from their home states. Even though the activity from Chennai, Delhi and Banglore is huge they account for less than one-fifth of the geo-tagged tweets talking about their home teams.

\subsection{Pearson Correlations}
In this section, we describe our analysis of linear dependencies between different sets of data using the Pearson product-moment correlation coefficient. We studied correlations between team popularity on social media with their brand values\footnote{http://iplt20wiki.com/ipl-brand-value-up-by-4-to-3-03-billion-ipl-2013/4312/} and IPL rankings.\footnote{http://www.iplt20.com/stats} Team popularity in this context means the amount a team is spoken of on social media as compared to others. Further, we studied correlations between team popularity on Facebook and Twitter. After having found correlations using tweet counts for each team, we went on to find similar correlations using the tweet count for the official IPL support tweet (I'm supporting {team name}! Who are you supporting? \#IPL \#{team abbreviation} via \@IPL) from the IPL website.\footnote{http://www.iplt20.com/} Table 5 shows the correlations we found between different datasets. We can clearly observe high linear dependence among a team's brand value, performance and its online presence. The correlation coefficient between the number of likes on a team's Facebook page and its brand value was calculated to be 0.92!

\begin{table}[ht]
\begin{tabular}{l l l l}
\\[0.1ex]\hline\hline
Variables &  Coefficient \\ [0.3ex] 
\hline \\ 
Facebook Page Likes, Team Brand Value  & 0.92  \\ [0.4ex] 
Facebook Page Likes, Tweets  & 0.83  \\ [0.4ex] 
Tweets, Team Brand Value & 0.79 \\ [0.4ex] 
Tweets, Followers on Twitter & 0.94 \\ [0.4ex]
Tweets, IPL Ranking & 0.70 \\ [0.4ex] 
IPL Website Tweet, Team Brand Value & 0.73 \\ [0.4ex]
IPL Website Tweet, IPL Ranking & 0.71 \\ [0.4ex] 
\hline 
\end{tabular}
\label{table:nonlin} 
\caption {Pearson Correlations} \label{tab:title} 
\end{table}

\vspace{-3 mm}

\subsection{Decay Analysis}

\begin{figure}[h!]	
                \centering
                \includegraphics[scale=0.29]{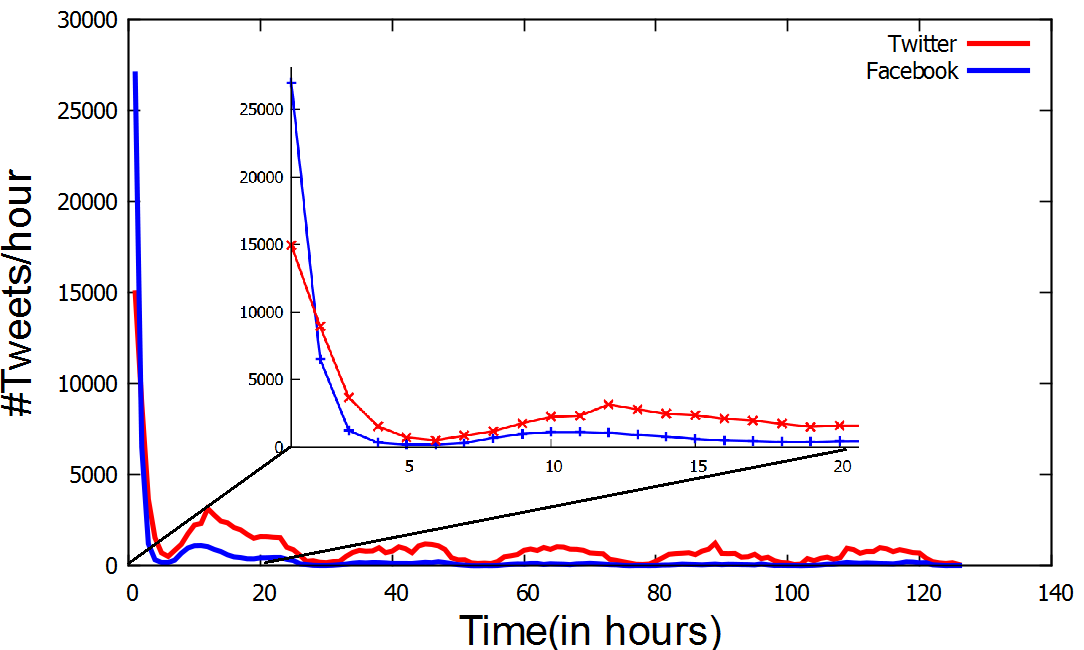}
                \caption{Fall in activity on Facebook and Twitter depicting similar decay patterns on both networks.}
                \label{fig:tiger}
\end{figure}

The plots shown in Figure 7 represent the fall in hourly tweet count after the finals of IPL between the teams Chennai Super Kings and Mumbai Indians. Prominent decay can be observed within the first 20 hours for both the OSNs. The plots clearly indicate a very similar decay pattern, which is also evident from the decay factor we calculated for IPL finals for both Facebook and Twitter. Applying the decay methodology (Section 4) on the plot for Facebook resulted in a decay factor of 0.1942 while for Twitter it was 0.2022. The decay factors are comparable as expected. Individual decay factors also indicate the steepness in fall of hourly tweets right after the finals.
It is also interesting to note that peak activity (the point in the plot from where decay starts) was higher for Facebook (27,005 posts) as compared to Twitter (14,988 tweets) while the decay became insignificant much earlier for Facebook than for Twitter. The decay threshold was taken as 99\% i.e the point beyond which the hourly count reduces to less than 1\% of the peak activity for a sustained period. Beyond the threshold, decay is considered insignificant for the net decay factor calculation.
This analysis could not be extended to individual matches due to overlap of decay and growth between two consecutive matches as in Figure 4(b).

\subsection{Sentiment Analysis}

We did a sentiment analysis on the content of tweets and Facebook posts using Linguistic Query and Word Count.\footnote{http://www.liwc.net/} We found that positive emotion in the text was higher than negative emotion (Figure 8). Words related to anxiety had a surprisingly low count, merely 0.33\%  and 0.31\% of the entire data for Twitter and Facebook respectively. Significant percentages were found for social words as well as words related to space and time. IPL'13 was the richest cricket tournament in the world\footnote{http://articles.timesofindia.indiatimes.com/2011-04-23/news/29466837\_1\_ipl-teams-nba-teams-indian-premier-league} and involved spot-fixing and money laundering incidents. Despite that, the percentage for money words is very low for both the networks. 

\begin{figure}[h!]
  \centering
    \includegraphics[width=0.47\textwidth]{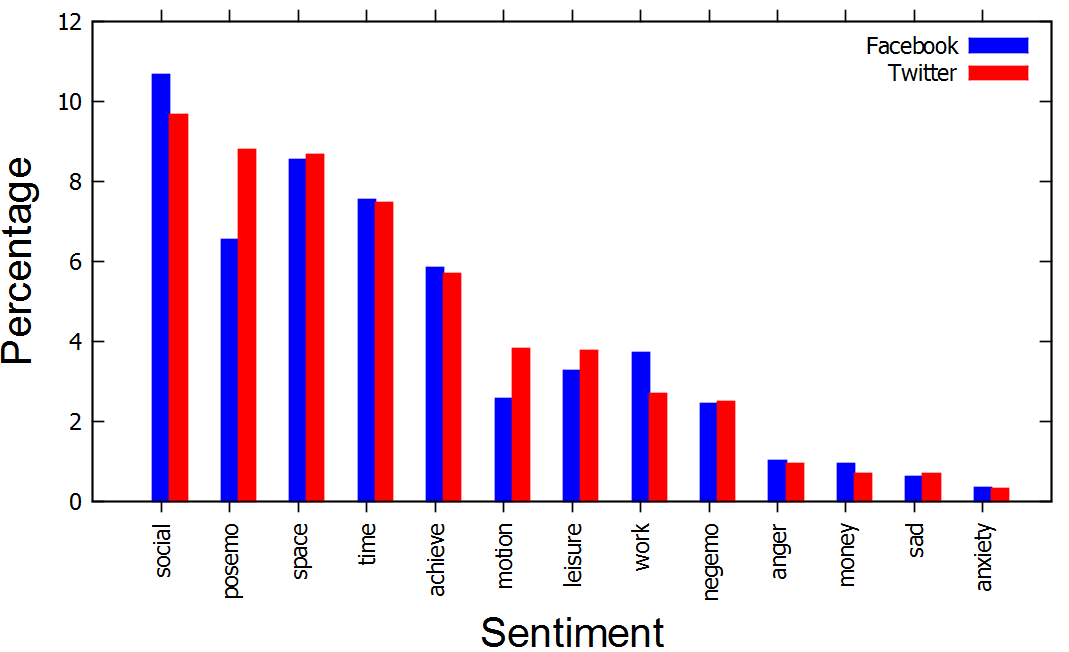}
   \caption{Text analysis of IPL dataset for both Twitter and Facebook showing high percentage of `social' and `positive-emotion' words.}
\end{figure}

We obtained the tweets and posts about spot-fixing and analysed them separately using LIWC. The results obtained for Twitter are shown in Figure 9. The percentage for anxiety words was less for this data too. Percentages for positive and negative emotions were less than the values obtained for the entire IPL data. We could also observe a higher percentage for money words being used in the context of matches being fixed. To understand these results in a better manner and  explain surprisingly low anxiety for the matches despite IPL's significant presence in both offline and online world, we analysed the tweet content manually. We found that IPL is more fun for people than anything else and they didn't care much about the matches being fixed. These results show that IPL serves more as a means of entertainment than anything else.

\begin{figure}[h!]
  \centering
    \includegraphics[width=0.48\textwidth]{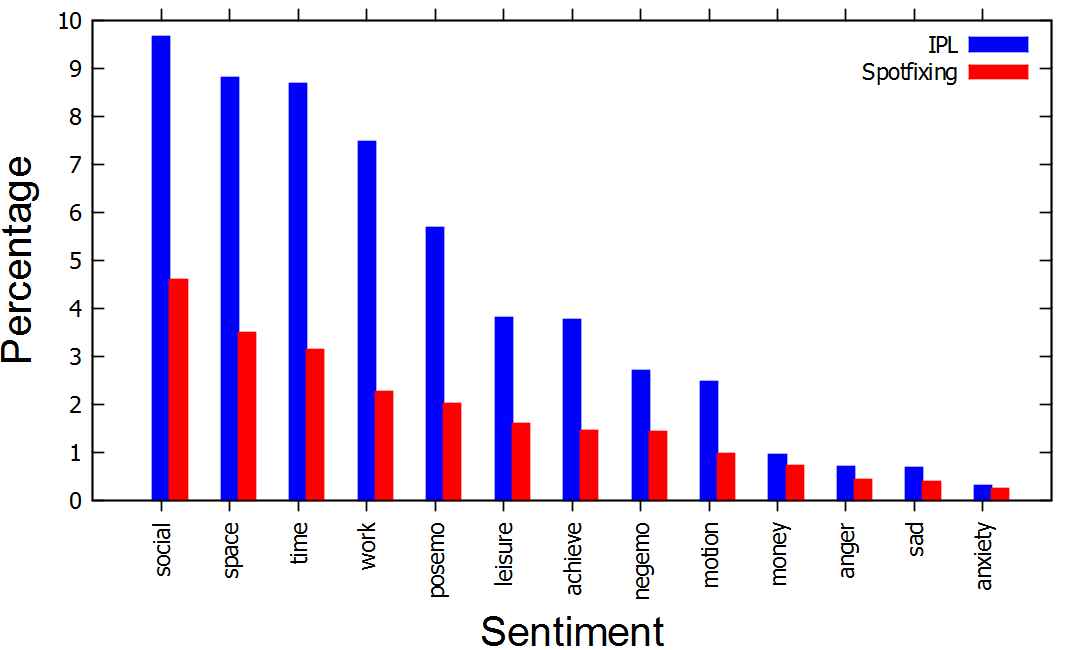}
   \caption{Text analysis of IPL and spot-fixing dataset for Twitter. The graph shows less percentage of  `positive-emotion' and `anxiety' words indicating that, despite huge fan-following and involvement, people view IPL as simple entertainment.}
\end{figure}

\section{Summary}

Online social media has become a platform to understand public opinion and sentiment. In this research work we analysed the online activity during the Indian Premier League. We studied 2,627,197 tweets and 774,186 Facebook posts from 485,533 unique users on Twitter and 
401,254 unique users on Facebook. The plots for hourly tweets vs. time clearly depicted the scheduled nature of the league. We analysed the geo-tagged tweets to understand the geographic distribution. Our results showed that 75.53\% of the geo-tagged tweets and 89.85\% of the geo-tagged posts were from India and of those the most significant contribution came from metropolitan suburbs and industrial belts. Decay analysis of the IPL activity yielded a net decay factor of 0.1942 for Facebook while for Twitter it was 0.2022. This clearly shows that the activity on both social networks died down similarly. We found strong correlations between the brand value of teams and how much they were talked of on Twitter (0.79) and Facebook (0.91). We also found that the correlations between popularity of different teams on both the OSNs were strong enough to say that they follow similar trends. Text analysis of both the spot-fixing dataset and the complete IPL dataset using LIWC showed that even after the spot-fixing scandal, people viewed IPL as simple entertainment and remained unaffected by the numerous allegations.

\vspace{2 mm}

To the best of our knowledge this is the first detailed analysis of an India-centric sports event on online social media. With more of such studies on the IPL and on other such events we can possibly find trends and draw parallels to better understand user behaviour in case of significant social and political events. We could also further expand this study using text analysis to find the number of supporters each team has and draw correlations similar to the ones drawn in this report. We are currently collecting data for IPL’14, and we plan to do a comparative analysis once IPL’14 is over.


\bibliographystyle{plain}
\bibliography{references}

\section{Appendix}
\begin{table}[ht]
\begin{tabular}{l l l l}
\hline\\[0.1ex]
`mumbai indians',`delhi daredevils',`rajasthan royals', \\ [0.4ex] 
`kings xi punjab',`pune warriors',`chennai superkings',\\ [0.4ex] 
`royal challengers bangalore',`kolkata knight riders',`RR', \\ [0.4ex] 
`IPL',`KKR',`DD',`KXIP',`PWI',`CSK',`RCB',`SH',\\ [0.4ex]
`sunrisers hyderabad',`PuneWarriors',`\#IPL',`\#IPL6',\\ [0.4ex]
`IPL6',`PepsiIPL',`KKRvsDD',`KorboLorboJeetbo', \\ [0.4ex] 
`IPLPulse',`Sunrisers',`\#PWI',`\#DD',`\#MI',`\#KKR',\\ [0.4ex]
`\#RCB',`\#RR',`\#PepsiIPL',`\#KKRvsDD',`\#KXIP',\\ [0.4ex] 
`\#Sunrisers',`\#PuneWarriors',`\#IPLPulse',\\ [0.4ex] 
`ipl2013',`\#KorboLorboJeetbo',`\#ipl2013'\\ [0.4ex] 
`apnemunde',`\#apnemunde'\\ [0.4ex]
\hline 
\end{tabular}
\caption {Data Collection Keywords} 
\end{table}

\begin{table}[!ht]
\begin{tabular}{l l l l}
\hline\\[0.1ex]
`meiyappan',`fix',`srinivasan',`bookie', \\ [0.4ex] 
`dara singh',`ramesh vyas',`pandurang kadam',\\ [0.4ex] 
`jiju janardhan',`ravi sawani',`rauf', \\ [0.4ex] 
`vikram agarwal',`ashwini agarwal',`ajay shirke',\\ [0.4ex]
`sanjay jagdale',`rajiv shukla', `jagmohan dalmiya',\\ [0.4ex]
`umesh goenka',`raj kundra',`sreesanth',`chandila', \\ [0.4ex] 
`ankeet chavan'\\ [0.4ex]
\hline 
\end{tabular} 
\caption {Keywords used for filtering spot-fixing data} 
\end{table}

\end{document}